# High performing additively manufactured bone scaffolds based on copper substituted diopside

Shumin Pang [a,*,1], Dongwei Wu [b,1], Franz Kamutzki [a], Jens Kurreck [b], Aleksander Gurlo [a], Dorian A.H. Hanaor [a,*]

[a] Technische Universität Berlin, Chair of Advanced Ceramic Materials, Straße des 17. Juni 135, 10623 Berlin, Germany
[b] Technische Universität Berlin, Chair of Applied Biochemistry, Gustav-Meyer-Allee 25, 13355 Berlin, Germany

## HIGHLIGHTS

- Robocasting of precipitation derived 0–3 at.% copper-doped diopside.
- Copper-doping improved mechanical properties.
- 1 at.% copper-doping showed favorable cell proliferation and *in vitro* angiogenic ability.
- Copper-doped diopside showed anti-*E. coli* activity, increasing with copper content.
- 1 at.% copper-doped diopside appears to be the optimal composition for scaffold material.

## GRAPHICAL ABSTRACT

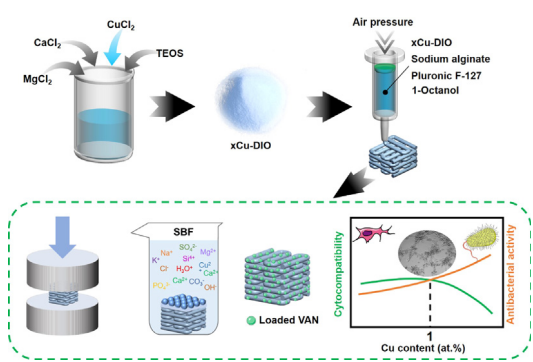

## ARTICLE INFO

## ABSTRACT

The inclusion of small amounts of copper is often reported to enhance the mechanical and biointegrative performance of bioceramics towards tissue engineering applications. In this work, 3D scaffolds were additively manufactured by robocasting of precipitation derived copper doped diopside. Compositions were chosen in which magnesium sites in diopside were substituted by copper up to 3 at.%. Microstructure, mechanical performance, bioactivity, biodegradability, drug release, biocompatibility, *in vitro* angiogenesis and antibacterial activity were studied. Results indicate that copper is incorporated in the diopside structure and improves materials' fracture toughness. Scaffolds with > 80% porosity exhibited compressive strengths exceeding that of cancellous bone. All compositions showed bioactivity and drug release functionalities. However, only samples with 0–1 at.% copper substitution showed favorable proliferation of osteogenic sarcoma cells, human umbilical vein endothelial cells and fibroblasts, while larger amounts of copper had cytotoxic behavior. *In vitro* angiogenesis was significantly enhanced by low levels of copper. Copper-containing materials showed anti-*Escherichia coli* activity, increasing with copper content. We show that across multiple indicators, copper substituted diopside of the composition $CaMg_{0.99}Cu_{0.01}Si_2O_6$, exhibits high performance as a synthetic bone substitute, comparing favorably with known bioceramics. These findings present a pathway for the enhancement of bioactivity and mechanical performance in printable bioceramics.

\* Corresponding authors.
E-mail address: Shumin.Pang@ceramics.tu-berlin.de (S. Pang).
[1] These authors contributed equally to this work.

## 1. Introduction

Over recent decades calcium silicate-based bioceramics have been the subject of diverse *in vitro* and *in vivo* studies as they combine properties of favorable bioactivity with osteogenic performance[1]. Recently, magnesium-containing silicates with good mechanical properties and biocompatibility have also attracted extensive attention [2]. Among them, diopside ($CaMgSi_2O_6$), a ceramic of the pyroxene family with a chain-like structure of corner-sharing $SiO_4$ tetrahedra flanked by $MgO_6$ and $CaO_6$ octahedra, has been identified as showing favorable mechanical and biological properties. In particular, fracture toughness in diopside (3.5 $MPa \cdot m^{1/2}$) [3] is significantly higher than that of comparable polycrystalline calcium silicate bioceramics, including the sorosilicate akermanite ($Ca_2MgSi_2O_7$) (1.83 $MPa \cdot m^{1/2}$) [4] and orthosilicate bredigite ($Ca_7MgSi_4O_{16}$) (1.57 $MPa \cdot m^{1/2}$) [5], while another chain silicate, the pyroxenoid wollastonite ($CaSiO_3$) has been found to have a fracture toughness of<1 $MPa \cdot m^{1/2}$ [6]. Diopside is known to exhibit the ability of forming osseointegrative hydroxyapatite (HAP, $Ca_{10}(PO_4)_6(OH)_2$) layers in simulated body fluid (SBF), and the release of magnesium and silicon from the diopside lattice has been shown to have positive effect on osteoblast proliferation, differentiation and bone growth [7]. Supporting this, an earlier study by Wu and Chang indeed identified the greater magnesium and silicon content as the key factors facilitating superior osteoblastic proliferation, apatite formation and biodegradation in diopside, relative to bredigite and akermanite ceramics [8]. Diopside bioceramics have been reported to exhibit intrinsic antibacterial activities [7]. Harnessing and enhancing the antiseptic bioactivity of diopside ceramics is a promising approach towards the mitigation of implant-related infections that may further aggravate bone defects and currently pose a major challenge in regenerative tissue engineering.

In the design of scaffold materials for effective bone tissue regeneration, bioceramic structures can be modified by incorporating appropriate ions into their structures to balance multiple bioactive functions, including biomineralization, osteogenesis, biodegradability and antibacterial activity. Copper is an essential trace element that plays an important role in maintaining the function of human cells, and is considered to be an effective inorganic antibacterial agent with low cytotoxicity. Studies have reported that both gram-negative and gram-positive bacteria can be effectively suppressed by copper [9]. The copper-induced enhancement of osteogenesis has been widely studied [10–12]. Copper ions are further known to stimulate the proliferation of vascular endothelial cells, thereby improving the angiogenic process, which constitutes a key aspect of bone repair [13]. Previous studies have shown that the copper ions promote the activity of a variety of enzymes and facilitate crosslinking of collagen and elastin in bones [11,14]. However, an excessive concentration of locally bioavailable copper may generate free radicals, which can interfere with bone metabolism [15]. The threshold concentration beyond which copper ions in a biomaterial may be considered to be cytotoxic depends on the type of adjacent cells and the release kinetics of copper from the host material [16]. Therefore, in the evaluation of copper-doped bioceramics, different copper concentrations and cell types need to be considered in the design of materials exhibiting enhanced osteogenesis and angiogenesis.

The control and customization of scaffold geometries across multiple scales are of high value in the processing of bioceramics towards patient-focused remedial medicine. In clinical practice, it is challenging to process porous scaffolds for bone tissue engineering with well-controlled shape and structure by traditional methods, such as those relying on space holder method [17]. Various emerging additive manufacturing (AM) technologies allow the production of scaffolds with complex structures and are increasingly being implemented in clinical treatments [18]. A particular method of AM, robocasting (also known as ceramic direct ink writing) is based on the direct extrusion of slurry. This AM technique allows the utilization of diverse powdered ceramic feedstocks in a robust and scalable approach. It overcomes the limitations of traditional scaffold manufacturing methods to adapt to complex bone defects, and can serve to precisely control the pore size and pore interconnectivity of the scaffolds [19,20].

In this work, inspired by the prospective performance of copper substituted diopside as a bioactive ceramic, we present a multifaceted investigation about the coprecipitation-based synthesis of such materials, their processing by robocasting, and the mechanical and biological performance of additively manufactured bone repair scaffolds. We aim to investigate and better understand the role that copper can play in modifying diopside with a view towards improving achievable combinations of functional properties in printable bioceramics.

## 2. Materials and methods

### 2.1. Synthesis of powders

Undoped and doped diopside powders were synthesized by the coprecipitation method. First, equimolar amounts of calcium chloride ($CaCl_2$, Carl-Roth, Karlsruhe, Germany, > 98%) and magnesium chloride hexahydrate ($MgCl_2 \cdot 6H_2O$, Carl-Roth, Karlsruhe, Germany, > 98%) were completely dissolved in ethanol and mixed with a double molar quantity of tetraethyl orthosilicate (($C_2H_5O)_4Si$, TEOS, VWR Chemicals) under agitation for 2 h. Aqueous ammonia (25 wt %, Merck) was added dropwise to the solution until pH reached 10 and the mixture was stirred overnight. Subsequently, the coprecipitated product was centrifuged and washed with ethanol twice. Finally, the product was dried at 80 °C overnight and calcined at 850 °C for 2 h in the ambient air. The synthesis of copper substituted diopside (together with diopside (DIO) collectively denoted in the following as xCu-DIO, where × indicates the atomic fraction of copper in respect to magnesium sites) powder was conducted similarly by replacing stoichiometric amount of $MgCl_2 \cdot 6H_2O$ with 0.5, 1, 2 or 3 at.% of copper (II) chloride ($CuCl_2$, Sigma-Aldrich, 99%) to produce $CaMg_{0.995}Cu_{0.005}Si_2O_6$ (named 0.5 Cu-DIO), $CaMg_{0.99}Cu_{0.01}Si_2O_6$ (1 Cu-DIO), $CaMg_{0.98}Cu_{0.02}Si_2O_6$ (2 Cu-DIO) and $CaMg_{0.97}Cu_{0.03}Si_2O_6$ (3 Cu-DIO), respectively.

### 2.2. Fabrication of scaffolds

xCu-DIO scaffolds were fabricated by robocasting to form a typical wood-pile structure, which was chosen as a representative scaffold morphology [21]. The synthetic powders were sieved to a particle size<63 μm. Sodium alginate (Alfa Aesar) and pluronic F-127 (Sigma Aldrich) were used as binders and 1-octanol (Merck) as the antifoaming agent. To prepare the printable slurry, 5.0 g synthetic powder, 0.3 g sodium alginate, 8.0 g pluronic F-127 solution (20 wt%) and 0.2 g 1-octanol were mixed using an overhead stirrer (RW16 basic, KIKA Labortechnik, Germany). The 3D printing system was based on a robotic deposition system modified from an Ultimaker 2 go printer (Ultimaker BV, Netherlands), reequipped with a simple air-pressure controlled syringe-style cartridge. Considering the differences in shrinkage of the scaffolds with diverse slurries, a series of models with relevant dimensions were designed and sliced into wood-pile structure using CURA software (version 4.6.2, Ultimaker). The G-code files from slicing were used to control the movement of the printer. Dispensing needles of different diameters (Vieweg, Germany) were used for printing so as to obtain the scaffolds with similar dimensions after sintering. Slur-ries were driven by a pneumatic setup with air pressures ranging from 2.5 to 3.5 bar. After printing, the green bodies were transferred to a desiccator for 2 days, which is essential to avoid cracking. Finally, the dry samples were sintered at 1250 °C for 3 h in the ambient air.

*2.3. Characterization of the powders and scaffolds*

The shrinkage of the scaffolds was calculated by measuring the length, width and height of the scaffolds before and after sintering. Particle sizes were obtained by analyzing water dispersed powder in a wet cell of an LS 13,320 (Beckman Coulter, USA) laser diffraction particle size analyzer. X-ray diffraction (XRD) patterns of xCu-DIO materials were obtained in a Bruker Axs D8 Advance X-ray diffractometer and using CuKα radiation, performed at 35 kV and 40 mA between 10 and 80° 2θ values in steps of 3 s/0.02°. The elemental composition and the copper oxidation state in the xCu-DIO materials were studied by X-ray photoelectron spectroscopy (XPS) on an Escalab 250 (Thermo Fisher Scientific, USA) with monochromatic Al K Alpha (1486.6 eV) at 2 kV and 1 μA. A surface diameter of 400 μm was analyzed with an acquisition time of 1203.5 s and XPS spectra calibrated to C1s at 284.8 eV. The microstructure of the scaffold surfaces was studied using scanning electron microscopy (SEM, LEO Gemini 1530, Carl, Zeiss, Germany). Energy-dispersive X-ray spectroscopy analysis (EDX) was performed with an EDX system (Thermo Fisher Scientific Inc., USA) coupled to an in-lens and secondary electron detector.

To measure the open porosity of the sintered scaffolds, the Archimedes method was used based on the following Eq. (1):

$$Porosity(\%) = \frac{W_1 - W_2}{W_1 - W_3} \times 100 \qquad (1)$$

where $W_1$ is the weight of the scaffolds with water, $W_2$ is the dry weight and $W_3$ is the wet weight of the scaffolds suspended in water. The measurement was conducted in six replicates.

*2.4. Characterization of mechanical properties*

We measured the compressive strength and modulus of xCu-DIO scaffolds (6 × 6 × 8 mm³) before and after 8 weeks soaking in Tris buffer solutions by compression between steel plates using a RetroLine mechanical testing machine (Zwick/Roell, Germany) equipped with a 10 kN load cell at a crosshead speed of 0.5 mm/min. The measurement was based on the ISO 17162:2014 standard and was conducted in six replicates. To quantify the fracture toughness ($K_{IC}$) and hardness (H) of xCu-DIO scaffolds, nanoindentation was performed with a Hysitron TI950 TriboIndenter (Bruker Corporation, Massachusetts, USA) equipped with a standard Berkovich diamond indenter tip. The testing was carried out in accordance with the ASTM E2546 standard. A maximum load of 10 mN with a holding time of 15 s and a loading and unloading rate of 1 mN/s were used. 50 indents were made for each sample in grid-wise arrangements of 5 × 10 indents. Indentation modulus (E) and H were calculated from the unloading part of the load-penetration curves by using the Oliver and Pharr method [22,23]. The fracture toughness in terms of $K_{IC}$ was calculated using the model of Anstis et al. [24] according to Eq. (2):

$$K_{IC} = 0.016\sqrt{\frac{E}{H}}\left(\frac{P}{c^{3/2}}\right) \qquad (2)$$

where P is the load, E is the modulus, H is the hardness, c is the crack length from the center of the indent to the crack tip.

*2.5. Drug release from xCu-DIO scaffolds*

The antibiotic medication Vancomycin Hydrochloride (VAN, Carl Roth, Karlsruhe, Germany) is used for treating a broad range of infections and sepsis including those occurring in bones and joints. For this reason, VAN was chosen as a model drug to evaluate the drug delivery properties of studied materials. First, VAN was dissolved in deionized water at a concentration of 1 mg/mL, and 0.5 g of scaffolds were immersed in 40 mL of this solution, and vibrated at 37 °C for 24 h. Then, the VAN-loaded samples were collected, washed gently, and dried at 37 °C in a vacuum oven for 24 h. The calibration curve for solutions containing different amounts of VAN was obtained using an ultraviolet–visible (UV–Vis) absorption spectrophotometer (UVmini-1240, Shimadzu, Japan) at a wavelength of 280 nm. To investigate VAN release, the VAN-loaded scaffolds were immersed in 2 mL phosphate buffered solution (PBS) under dynamic conditions at 37 °C. At a certain time, 1 mL supernatant was collected and replaced with fresh PBS. Finally, the release amount of VAN was measured at 280 nm using the spectrophotometer, conducted in three replicates.

*2.6. In vitro bioactivity and biodegradability evaluation*

The dissolution of bioceramics and the formation of a surface layer of precipitated HAP are key indicators of osseointegration. To evaluate the effect of copper doping on apatite-formation ability, we prepared SBF containing ion concentrations similar to human blood plasma according to previous literature [25]. The scaffolds were soaked in SBF at 37 °C for 28 days. During immersion, the changes in pH value of the solutions were measured. After 28 days of immersion, scaffolds were dried at 60 °C for 24 h and then studied by XRD, SEM and EDX.

The degradation rate of scaffolds was investigated in Tris buffer with an initial pH of 7.25 at 37 °C. The buffer solution was changed every 3 days. After 8 weeks, the scaffolds were removed from the buffer solution and dried at 60 °C for 24 h. Finally, the weight loss of the scaffolds was calculated as follows:

$$Weight\ loss(\%) = \frac{W - W_1}{W} \times 100 \qquad (3)$$

where W is the weight of the scaffold before soaking and $W_1$ is the weight after soaking.

*2.7. Cytocompatibility*

Cell culture

Human osteogenic sarcoma cells (Saos-2 cells, Charité, Department of Periodontology, Oral medicine and Oral surgery), human umbilical vein endothelial cells (HUVECs; ATCC CRL-1730), and primary human lung fibroblasts (ATCC PCS-201–013) were chosen to study the effects of xCu-DIO scaffolds on cell viability and proliferation. The Saos-2 cells, HUVECs and fibroblasts were cultured in an incubator (37 °C, 5% $CO_2$) with a humidified atmosphere. Complete media were used for cell culture and were refreshed every other day. Specifically, for Saos-2 cells: Dulbecco's Modified Eagle medium (DMEM, High Glucose, Biowest) with 15% fetal bovine serum (FBS), 2 mM L-Glutamine; for HUVECs: MCDB 131 medium (Gibco, Thermo Fisher Scientific) with 10% FBS, 2 mM L-Glutamine; for fibroblasts: DMEM with 10% FBS and 2 mM L-Glutamine.

*2.8. Cell viability assay*

We evaluated cell viability following exposure to the extracts of the scaffolds using 2,3-bis(2-methoxy-4-nitro-5-sulfophenyl)–2H-tetrazolium-5-carboxanilide sodium salt (XTT sodium salt, Alfa

Aesar). Cells were seeded in 96-well culture plates, and after cell attachment, the supernatant from each well was replaced with 100 μL test sample extract. The cells were further cultured for 1, 3 and 7 days. Then, the XTT solution and phenazine methosulfate (PMS, AppliChem, Darmstadt, Germany) was added (final concentration: 0.3 mg/mL XTT and 2.3 μg/mL PMS) and incubated for 4 h. The absorbance of the supernatant at 450 nm was measured using a microplate reader (Tecan Austria GmbH, Austria). Fresh medium under the same conditions was also tested as background. Each test was performed with six parallel samples.

A fluorescent live/dead assay kit (L3224; Thermo Fisher Scientific) was used to stain the aforementioned cells to assess their viability on scaffolds. Briefly, after the cells were seeded on each scaffold for 24 h, the scaffolds were rinsed with PBS and incubated in a working solution (containing 2 μM ethidium homodimer-1 and 2 μM calcein AM) at 37 °C with 5% $CO_2$ for 15 min. Subsequently, fluorescence was observed by fluorescence microscopy (Observer Z1, Zeiss), conducted in three replicates. Live cells were presented in green while dead cells were shown as red.

### 2.9. In vitro angiogenesis

In order to evaluate the effects of Cu-DIO on angiogenic activity *in vitro*, a tube formation assay was conducted using matrigel (Corning Matrigel Matrix) [26]. Briefly, 96-well culture plates were coated with 50 μL matrigel. Then, HUVECs (1 × $10^4$ cells per well) were cultured in Cu-DIO extracts with matrigel underneath for 4 and 8 h, and HUVECs cultured on matrigel were set as control. Cu-DIO extracts were prepared by immersing 1 g of material in 5 mL serum-free DMEM and incubating at 37 °C for 24 h. The supernatant was then collected and sterilized through a filter (Millipore, 0.22 mm). At each time point, the capillary tube branch points formed by HUVECs were photographed using a microscope (Observer Z1, Zeiss) in five random microscopic fields per well, and counted.

### 2.10. Antibacterial activity

We examined the antibacterial activity of xCu-DIO scaffolds towards *Escherichia coli* (*E. coli*, ATCC 25922) bacteria following liquid medium microdilution techniques and agar diffusion assay. For liquid medium microdilution, 0.3 g scaffold specimens were immersed in bacterial suspensions known to contain ∼ 1 × $10^5$ CFU/mL for a period of 24 h. The bacterial solution without any treatment was set as a control. The absorbance values of all tested solutions were read at 600 nm using a UV–Vis spectrophotometer. Each test was performed with six parallel samples. The percentage of bacterial inhibition was calculated by the following Eq. [27]:

$$Inhibition(\%) = \frac{I_c - I_s}{I_c} \times 100 \quad (4)$$

where $I_C$ and $I_S$ are the absorbances of the control bacterial suspension and an experimental bacterial suspension containing different xCu-DIO scaffolds, respectively.

Agar diffusion assays were further used to assess the antibacterial activity of the scaffolds. For this, a 100 μL *E. coli* suspension (1 × $10^6$ CFU/mL) was seeded on Luria-Bertani (LB) agar medium plate. Scaffold extracts were prepared by immersing 0.3 g specimens of scaffold materials in LB medium and incubating at 37 °C for 24 h. Then, sterilized 10-mm filter paper discs were immersed in this LB medium. Filter paper discs that had been immersed in sterile deionized water were used as a control. Subsequently, these filter paper discs were placed on the surface of the *E. coli* seeded agar medium and then the plate was cultured for 24 h at 37 °C. Finally, the inhibition zone was determined on the basis of images obtained by a digital camera. Each test was performed with three parallel samples.

### 2.11. Statistical analysis

In this study, all data were expressed as mean value ± standard deviation (SD). Student's *t*-test and analysis of variance (ANOVA) were used to analyze differences between experimental groups. Results are considered significantly different as *$p < 0.05$ and **$p < 0.01$.

## 3. Results

### 3.1. Structural characterization of scaffolds

Fig. 1a shows photographs of a representative diopside scaffold with a periodic woodpile structure, where the filament spacing distance is approximately 350 μm. Copper doping caused an increasing blue coloration of the scaffolds (Fig. 1b). The shrinkage of xCu-DIO scaffolds after sintering was between 31% and 36% (Fig. 1c). As shrinkage depends on the particle size and green body density, and the average particle size of xCu-DIO powders increased with copper doping (Fig. 1d); increasing copper content results in smaller shrinkage of scaffolds [28].

XRD patterns of sintered scaffolds are presented in Fig. 2a. The XRD patterns of the undoped and doped materials indicate solely the characteristic peaks of diopside, $CaMgSi_2O_6$ (PDF # 72–1497), with no significant presence of other crystalline phases. It can be seen that copper doping slightly enhanced the intensity of the diffraction peaks which are related to crystallinity and crystallite size. This may be that copper acted as a flux, lowering the sintering temperature of diopside; therefore, for the given sintering temperature of 1250 °C the crystallinity and grain size of the copper-doped samples was increased [29,30].

The overview XPS spectra of xCu-DIO samples are shown in Fig. 2b, displaying the peaks related to the diopside composition (Mg, Ca, Si and O). Binding energies are consistent with the expected values [31]. The Cu 2p photoelectron spectrum (Fig. 2c) was used to estimate the oxidation state of copper ions in the diopside. The peaks close to 952 eV and 932 eV correspond to Cu 2p1/2 and Cu 2p3/2 spin–orbit splitting, respectively. With the increase of copper doping, the binding energy at 2p3/2 decreases, which was 933.1 eV, 933.0 eV, 932.8 eV and 932.7 eV respectively. The broad peak at 943 eV is a characteristic satellite of $Cu^{2+}$ [32]. It can be seen from Fig. 2c that the Cu 2p3/2 spectra of all samples were symmetrical and only fitted to one Gauss-Lorentz peak. Therefore, it can be concluded that copper in all samples mainly existed in the form of $Cu^{2+}$ [33].

Fig. 3 presents SEM micrographs of sintered diopside ceramic scaffolds with different copper content. All the scaffolds had a typical sintered porous morphology consisting of equiaxed submicron grains, which grew larger with increasing copper content. This result indicates that copper promoted the sinterability of diopside ceramics, which can be used to reduce the time and energy consumption in ceramic production. The elemental composition of diopside and 1 Cu-DIO was further confirmed by the EDX analysis.

### 3.2. Mechanical testing

Multi-scale porosity and good mechanical strength are instrumental in scaffolds for bone tissue engineering. Table 1 lists the porosity and mechanical properties of human bone and the different scaffolds fabricated here. According to the Archimedes' method, all samples were measured to have a porosity of about

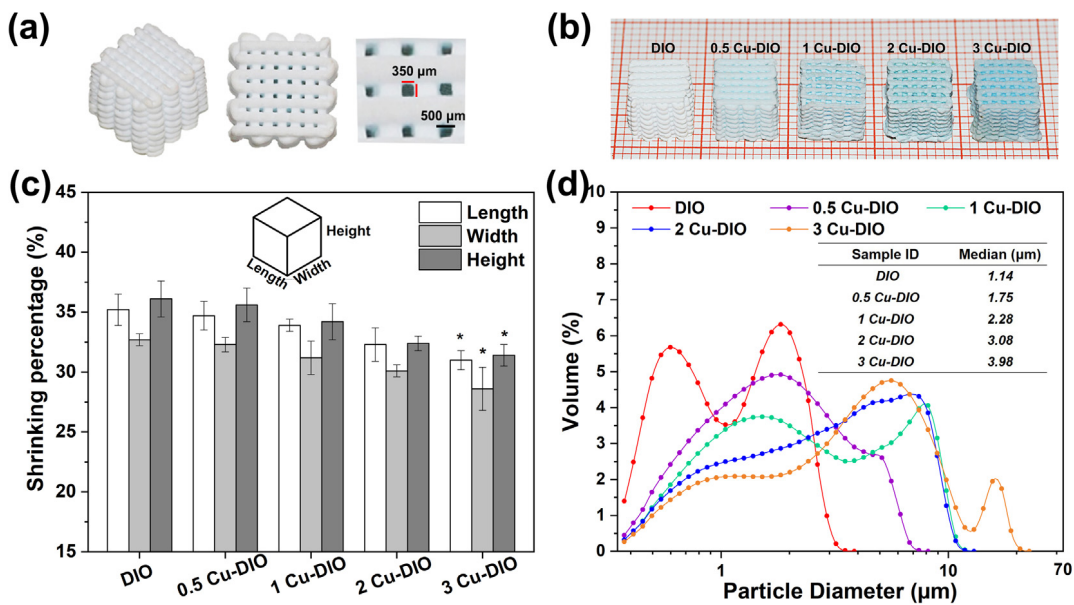

**Fig. 1.** Characterization of xCu-DIO scaffolds. (a) Photographs of a representative diopside scaffold, (b) photograph of 3D printed diopside scaffolds with different levels of copper doping, (c) linear (length, width, height) shrinkage (*p < 0.05, compared with DIO group, respectively. n = 12), and (d) particle size distribution of xCu-DIO powders after sintering (n = 5).

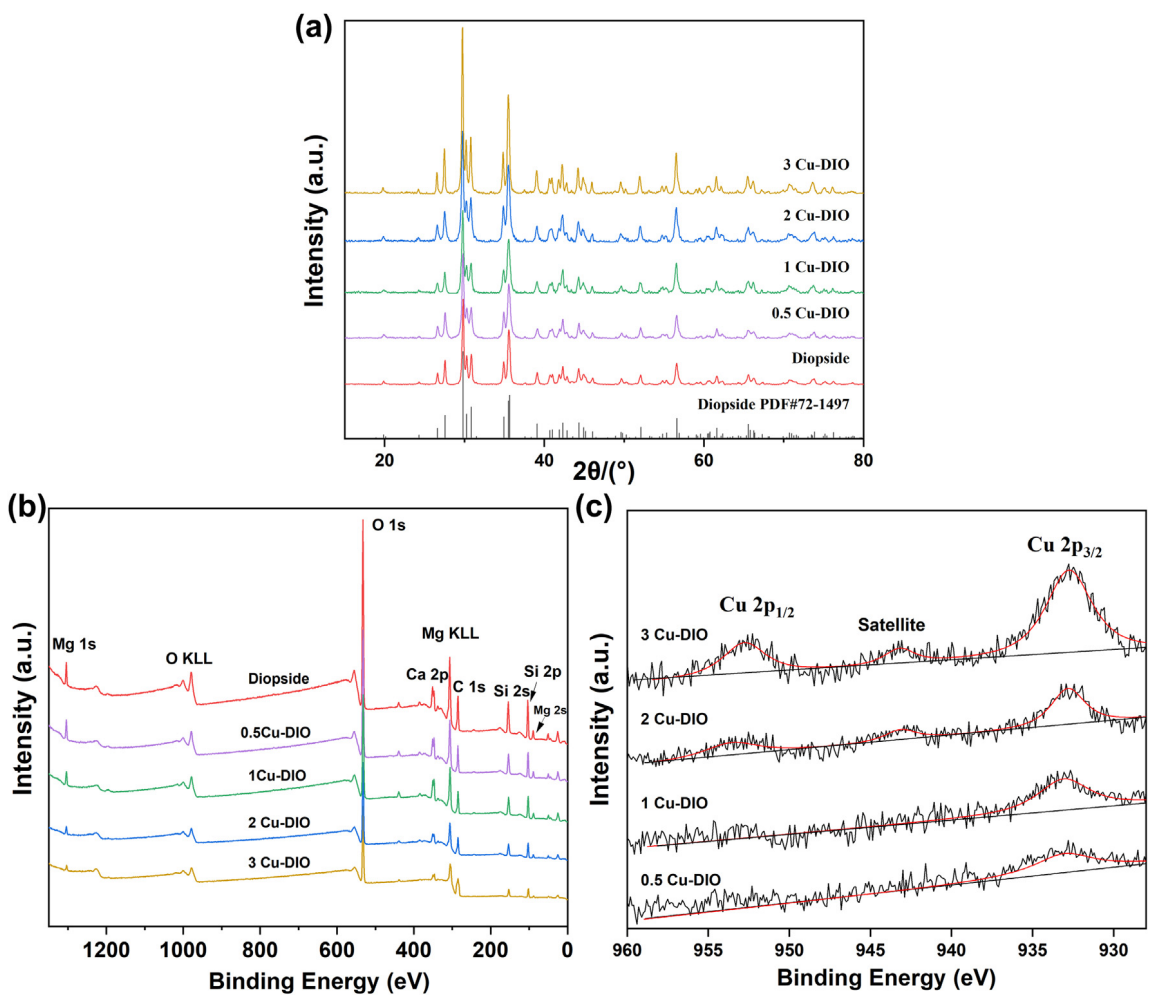

**Fig. 2.** XRD and XPS analysis of xCu-DIO samples after sintering at 1250 °C. (a) XRD patterns; Standard spectra of diopside is provided below, (b) the overview XPS spectra and (c) core level Cu 2p spectra of 0.5 to 3 Cu-DIO samples.

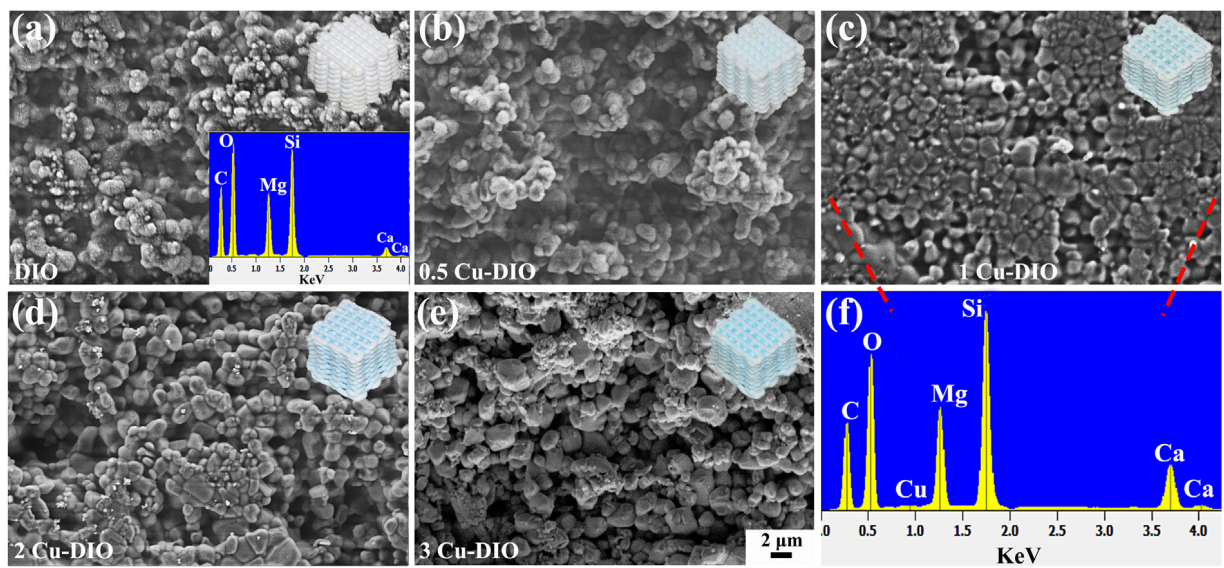

**Fig. 3.** Morphology and chemical analysis of xCu-DIO scaffolds. (a) DIO micrograph and its corresponding EDS spectrum, (b) 0.5 Cu-DIO, (c) 1 Cu-DIO, (d) 2 Cu-DIO, (e) 3 Cu-DIO, and (f) EDX spectrum of 1 Cu-DIO sample.

**Table 1**
Porosity, compressive strength, compressive modulus, fracture toughness and hardness of xCu-DIO scaffold, human bone and common biomaterials.

| Sample | Porosity (%) | Compressive strength (MPa) | Compressive modulus (GPa) | Fracture toughness (MPa·m$^{1/2}$) | Hardness (GPa) |
|---|---|---|---|---|---|
| **Cancellous Bone** [34] | 30–90 | 4–12 | 0.1–0.5 | 0.1–0.8 | 0.4–0.6 |
| **Cortical Bone** [34] | 5–13 | 130–180 | 12–18 | 2–12 | 0.6–0.7 |
| **Bioglass (3DP)** [35] | 60 | 16 | 0.15 | – | – |
| **β-TCP(3DP)** [36] | 56 | 16 | – | – | – |
| **DIO** | 80.7 ± 1.1 | 27.83 ± 2.16 | 0.188 ± 0.030 | 3.13 ± 0.12 | 11.07 ± 1.03 |
| **0.5 Cu-DIO** | 81.3 ± 1.3 | 27.24 ± 4.49 | 0.190 ± 0.042 | – | – |
| **1 Cu-DIO** | 80.1 ± 1.5 | 29.16 ± 3.43 | 0.232 ± 0.035 | 3.16 ± 0.24 | 11.24 ± 0.72 |
| **2 Cu-DIO** | 81.5 ± 0.8 | 29.77 ± 3.81 | 0.239 ± 0.012 | 3.37 ± 0.10 | 11.44 ± 0.81 |
| **3 Cu-DIO** | 81.1 ± 1.2 | 31.43 ± 2.46 | 0.265 ± 0.038 | 3.43 ± 0.18 | 12.32 ± 0.86 |

3DP: 3D Printing.

80%, which is conducive for the inward growth and regeneration of tissue [34].

The compressive strength of scaffolds shown in Table 1 is the consequence of multiple factors, including the scaffold geometry. Results show that for the woodpile scaffolds produced here, copper doping led to an increase in the compressive strength from 27.83 ± 2.16 GPa to 31.43 ± 2.46 GPa and in the compressive modulus from 0.188 ± 0.030 GPa to 0.265 ± 0.038 GPa. To ascertain the fracture toughness and hardness of these novel bioceramics, nanoindentation tests were performed to evaluate the $K_{IC}$ and $H$ of the scaffolds. With the increase of copper content, the fracture toughness $K_{IC}$ of the diopside materials gradually and monotonically increased from 3.13 ± 0.12 to 3.43 ± 0.18 MPa·m$^{1/2}$. Similarly, the hardness of the materials increased too with increasing copper doping bringing a monotonic increase of H from 11.07 ± 1.03 to 12.32 ± 0.86 GPa. The mechanical properties of the scaffolds studied were between the upper limit of cancellous bone and the lower limit of cortical bone.

### 3.3. Drug release from xCu-DIO scaffolds

The cumulative release of the VAN from xCu-DIO scaffolds is presented in Fig. 4. Apparently, all the samples had rapid drug release within the first 24 h, due to the dissolution of surface-bound VAN. Subsequently, the release rate decreased significantly with time, and the xCu-DIO scaffolds maintained continuous release after 360 h (total release ∼ 93%). Therefore, all scaffolds were capable of adsorbing drugs and subsequently maintaining sustained release.

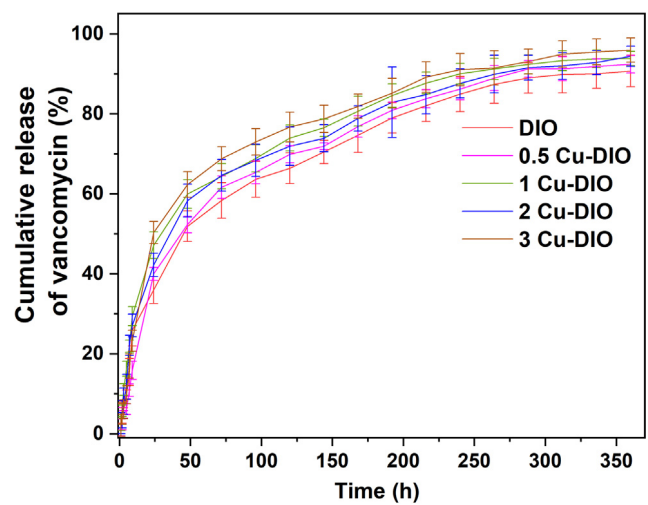

**Fig. 4.** The cumulative VAN release from xCu-DIO scaffolds for different time periods (n = 3).

### 3.4. In vitro bioactivity and biodegradability

*In vitro* bioactivity

Immersion in an SBF solution can be used as an *in vitro* test method to study the formation of the apatite layer on the implant surface to predict osseointegration *in vivo*. Fig. 5a depicts the change in pH value of SBF during 28 days of soaking for various scaffolds. The changes in pH indicate the occurrence of ionic exchanges between the scaffolds and the solution, and the onset of apatite deposition [37].

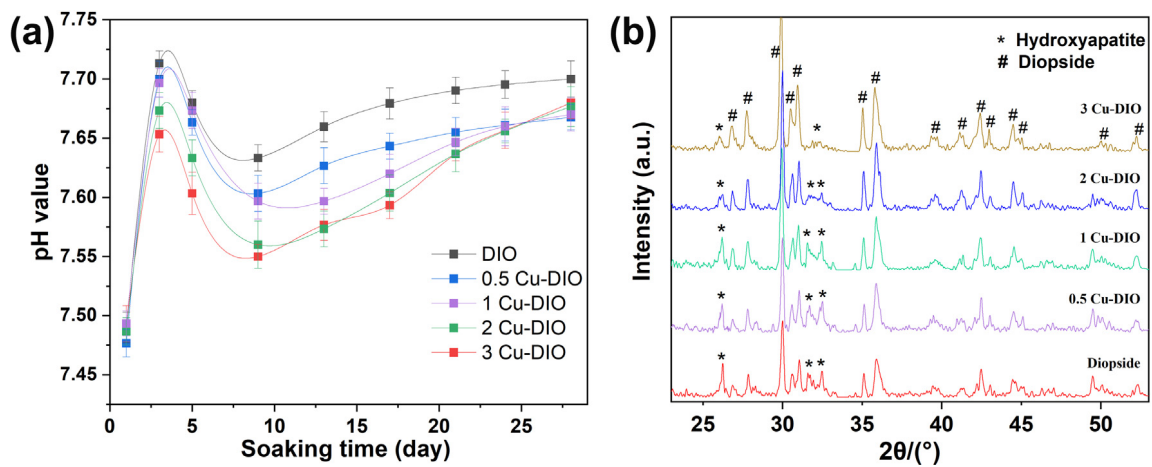

**Fig. 5.** (a) Changes in pH after immersion in SBF for different time periods, and (b) XRD of xCu-DIO scaffolds after 28 days of immersion in SBF.

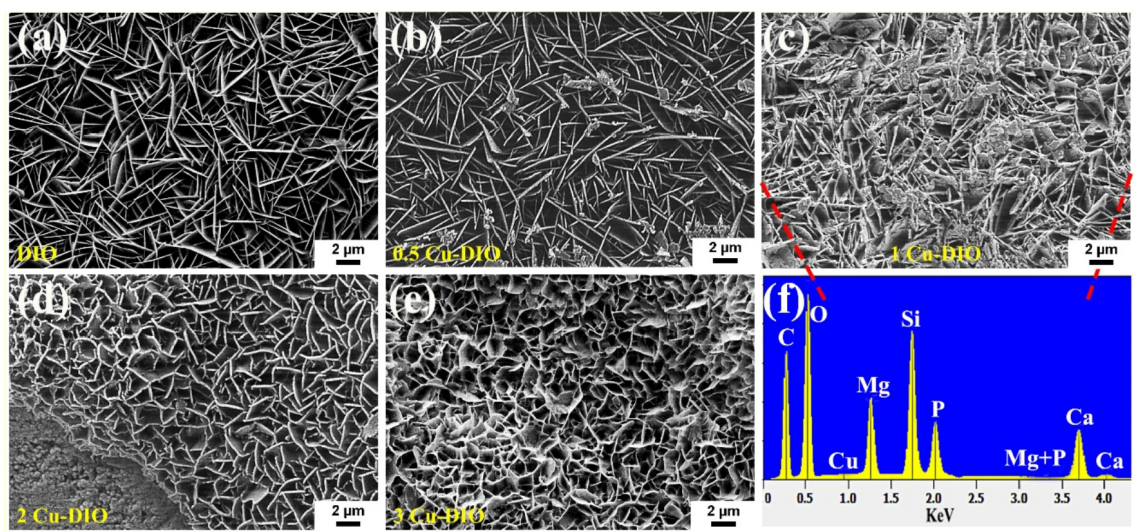

**Fig. 6.** SEM micrographs of xCu-DIO scaffolds after immersion in SBF for 28 days. (a) DIO, (b) 0.5 Cu-DIO, (c) 1 Cu-DIO, (d) 2 Cu-DIO, (e) 3 Cu-DIO, and (f) EDX spectrum of 1 Cu-DIO sample.

Fig. 5b shows the XRD patterns of scaffolds after immersion in the SBF for 28 days. The peaks at 25.82°, 31.77° and 32.49°, are attributed to the (0 0 2), (2 1 1) and (3 0 0) planes of HAP (PDF#74-0565), and can be found on all scaffolds following 28 days' immersion. This finding shows the precipitation of nanocrystalline HAP, indicating the osseointegrative bioactivity of these materials, which was further studied by SEM and EDX.

SEM micrographs of different scaffold surfaces following SBF immersion are shown in Fig. 6. The formation of precipitated HAP in such morphology is a typical characteristic of bioactive ceramics [38,39]. The presence of a strong phosphorus signal in the EDX spectrum of the 1 Cu-DIO scaffold, shown in Fig. 6f, alongside peaks of Ca, Mg and Si, supports the conclusion that a HAP layer had formed on the underlying diopside.

### 3.5. Degradation behavior and mechanical properties in vitro

Effective bone tissue engineering scaffolds should be bioresorbable, i.e., they progressively dissolve and are used to replace bone tissue without toxicity or rejection [40]. The weight loss of the xCu-DIO scaffolds following 8 weeks of soaking in Tris buffer solution are shown in Table 2. It can be seen that the incorporation of copper ions into the diopside scaffold had a certain effect on the degradation behavior, with an increase in weight loss from 9.3% (DIO) to 11.2% (3 Cu-DIO). The discrepancies between the ionic radius of $Cu^{2+}$ (0.73 Å, CN = 6) and $Mg^{2+}$ (0.72 Å, CN = 6) ions naturally induce degree lattice disorder in the form of point defects, which is generally expected to improve the degradation behavior of the scaffold [41,42]. The compressive strength of scaffolds studied here decreased from 27.83 to 31.43 MPa to 15.52–17.48 MPa after 8 weeks of immersion in Tris buffer. This is a natural consequence of dissolution during the immersion process leading to grain boundary weakening [43].

**Table 2**
Weight loss and compressive strength of xCu-DIO scaffolds after immersion in Tris buffers for 8 weeks.

| Sample ID | Weight loss (%) | Compressive strength (MPa) |
|---|---|---|
| **DIO** | 9.3 ± 1.2 | 15.52 ± 1.35 |
| **0.5 Cu-DIO** | 10.7 ± 1.9 | 14.71 ± 3.13 |
| **1 Cu-DIO** | 10.4 ± 1.5 | 16.36 ± 1.69 |
| **2 Cu-DIO** | 10.6 ± 1.2 | 17.12 ± 1.85 |
| **3 Cu-DIO** | 11.2 ± 1.3 | 17.48 ± 2.21 |

### 3.6. Cytocompatibility

It is well known that copper ions are cytotoxic above a certain concentration, and the cytotoxic effects of copper ions on different cells vary [44]. Based on this, the cytotoxicity of xCu-DIO towards different relevant cell types must be studied to design bone scaffold materials that facilitate osteogenesis and angiogenesis. Osteoblasts are essential for bone growth, function, repair, and maintenance. Furthermore, bone growth is largely dependent on vascular networks for the transport of nutrients/metabolites, and HUVECs play a crucial role in the formation of new vascular networks. In addition to endothelial cells, fibroblasts are also involved in angiogenesis [45]. Motivated by this, to assess cytocompatibility we examined the influence of copper content in xCu-DIO scaffold materials on the viability of Saos-2 cells, HUVECs and fibroblasts.

Fig. 7 shows the effect of xCu-DIO scaffold material ion extract on the proliferation of three cell types. The three cell types showed similar proliferation trends. Scaffolds with 0–1 at.% copper substitution at magnesium sites were found to promote the proliferation of Saos-2 cells, HUVECs and fibroblasts over 7 days of culture relative to the control measurement. However, diopside materials with higher copper contents, namely those with 2 or 3 at.% copper substitution, were found to show clear cytotoxicity towards Saos-2 cells, HUVECs and fibroblasts ($p < 0.01$). This influence was dose-dependent.

We further studied the growth of these three cell types on diopside scaffolds by fluorescence microscopy after live-dead staining (Fig. 8). Adhesion of Saos-2 cells, HUVECs and fibroblasts were particularly strong to the 0.5 and 1 Cu-DIO scaffolds. The results of cell proliferation and cell vitality tests were consistent, so we infer that 0.5 and 1 Cu-DIO has beneficial osteogenic and angiogenic effects on Saos-2 cells, HUVECs and fibroblasts.

### 3.7. In vitro angiogenesis

As the growth of blood vessels, through processes of branching and growth of existing vessels, known as angiogenesis, is a key aspect of bone defect healing, effective bone tissue engineering materials should promote this process, which is known as angiogenic bioactivity. In this work, the angiogenic bioactivity of Cu-DIO extracts was assessed *in vitro*. After 4 h of incubation on matrigel, HUVECs self-assembled and formed branch points in all culture media (Fig. 9a), while HUVECs grown in the 1 Cu-DIO extract showed not only interconnected branch points (blue point in Fig. 9a) and mesh-like circles (yellow line in Fig. 9a), but also formed significantly more tube-like parallel cell lines (tubes, red line in Fig. 9a), representing the typical morphology of late angiogenesis [46]. Compared to the control group, HUVECs cultured in 0.5 and 1 Cu-DIO extracts media formed a more complex geometry with a greater number of branch points, indicating a more effective angiogenesis. However, HUVECs in the 2–3 Cu-DIO extract group formed fewer branch points and tubes, indicating a slower angiogenesis due to the observed cytotoxicity imparted by higher copper content. After 8 h of incubation (Fig. 9b), sparser mesh-like structures were observed in all groups, while HUVECs cultured in 1 Cu-DIO extract still maintained extensive branch points and tubes, although a slightly reduced number of branch points was seen in comparison with the measurement following 4 h of incubation (Fig. 9c) [26].

### 3.8. Antibacterial activity

The antibacterial activity of xCu-DIO scaffolds towards *E. coli* was evaluated after culturing for 24 h. As shown in Fig. 10a (left Y-axis), the antibacterial inhibition rate of the pristine diopside sample was around 50%, which is consistent with reports of the antibacterial activity of this phase by Choudhary et al. [47]. Compared to a pristine diopside sample, Cu-DIO scaffolds show enhanced antibacterial activity, which increased with copper content. This trend is further confirmed by Fig. 10b, which shows the spatial inhibition of *E. coli* growth in the sterilized filter paper supported scaffold extracts, following a standard agar diffusion test. The quantitative results of the inhibition zone are shown in Fig. 10a (right Y-axis). This also indicates that for Cu-DIO samples the antibacterial activity was dose-dependent.

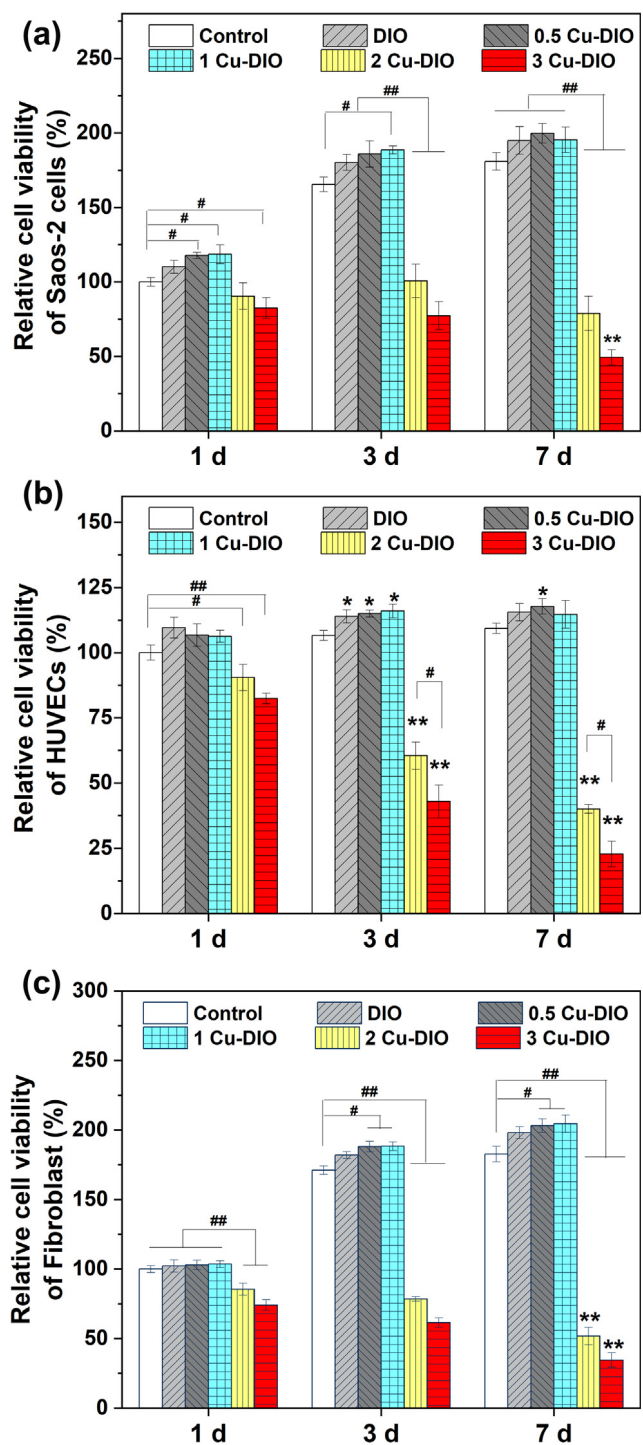

Fig. 7. *In vitro* cell viability after 1, 3 and 7 days exposure to scaffold extracts, as determined by XTT assay. (a) Saos-2 cells, (b) HUVECs, and (c) fibroblasts. (n = 6, *$P < 0.05$ and **$P < 0.01$, compared with 1 day group, respectively. #$P < 0.05$ and ##$P < 0.01$, as pointed out in the figure).

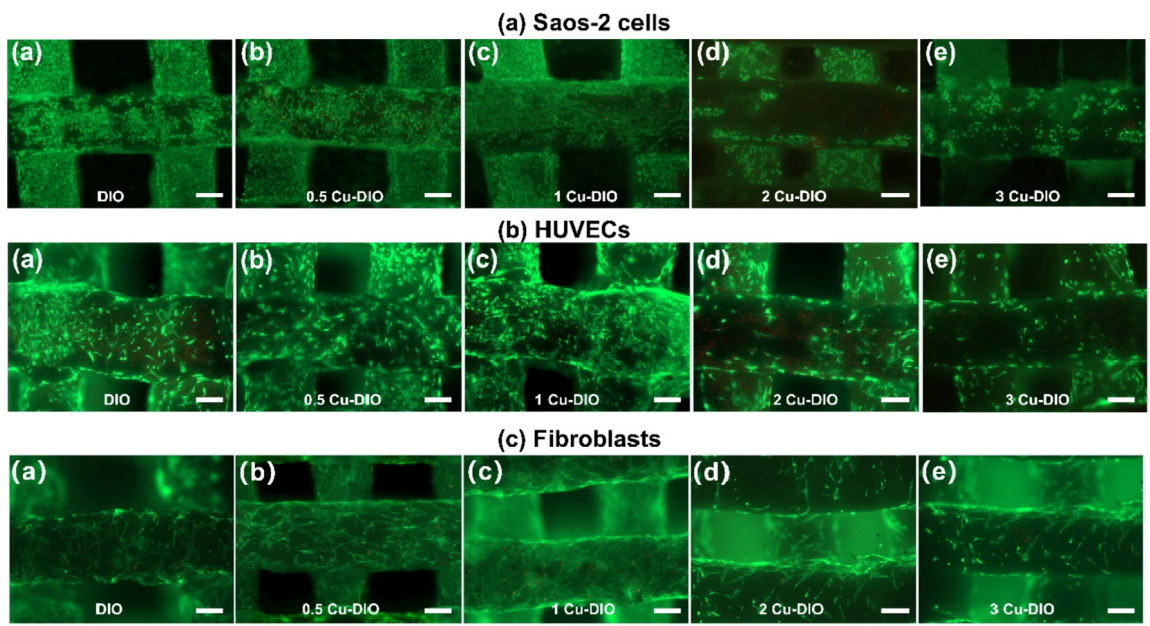

**Fig. 8.** Fluorescence images of Live/Dead staining of (a) Saos-2 cells, (b) HUVECs and (c) fibroblasts cultured on the xCu-DIO scaffold surface for 3 days. Live cells were stained to be green and dead cells were in red. Bar = 200 μm. (For interpretation of the references to colour in this figure legend, the reader is referred to the web version of this article.)

## 4. Discussion

In this work, we examined the fabrication and performance of additively manufactured bioceramic scaffolds based on copper substituted diopside. The design and processing of these materials were done with a view towards bone tissue engineering scaffolds, an important biomedical application that imposes multiple material challenges. XRD results suggest that copper was incorporated into the diopside structure rather than forming secondary oxide phases (Fig. 2a). Nevertheless, grain boundary segregation and the formation of amorphous nanoscale copper-rich phases cannot be ruled out based on these results. Since the ionic radius of $Cu^{2+}$ (0.73 Å, CN = 6) is close to that of $Mg^{2+}$ (0.72 Å, CN = 6), the substitution of copper at magnesium sites at the low doping levels studied here is not expected to bring about significant lattice distortion. Cu 2p X-ray photoelectron spectra confirm $Cu^{2+}$ oxidation state in all samples studied (Fig. 2c). Doping with copper increases the homologous temperature during sintering of diopside [30], and larger grains are formed in sintered copper-doped samples as seen by SEM (Fig. 3). The $Mg_{2-x}Cu_xSiO_4$ (x = 0–0.4) ceramics prepared by Lai, Y et al. showed an increase in grain size with increasing $Cu^{2+}$ ion content at 1250 °C sintering, which is consistent with our results [48].

### 4.1. Mechanical properties

Porosity and mechanical strength are key parameters of bone tissue engineering scaffolds. The porosity threshold of bioscaffold materials is about 60%, below which it does not provide sufficient space for cell adhesion and inward bone growth [49]. As with most other AM techniques, porosity can be easily modified by adjusting printing parameters. In our research, the scaffolds had reasonably high porosity (~80%), and the mechanical properties improved with a higher copper content (Table 1). The scaffolds fabricated here, in simple woodpile morphologies showed good levels of compressive strength compared to 3D printed bioglass [35] and β-TCP [50]. As with porosity, the compressive strength of these scaffolds can be adjusted via a plethora of possible processing and scaffold design modifications. The comparably good compressive strength of the scaffolds indicates the suitability of robocast diopside scaffolds towards the repair of injuries associated with cancellous bone. Moreover, the strength of xCu-DIO scaffolds was still sufficient for the repair of damaged cancellous bone even after soaking in Tris buffer for 8 weeks (Table 2). Fracture toughness values obtained by the applied nanoindentation method are similar to values for diopside reported elsewhere [4]. With increasing copper content, higher values of compressive strength and fracture toughness were obtained. This is most likely the result of copper serving as a flux. For a given processing temperature, copper-doped materials exhibit a denser structure with fewer crack-initiating voids, which further improves compressive strength and fracture toughness. However, the sintering of polycrystalline bioceramics at higher temperatures tends to produce a larger grain size, and thus slower degradation kinetics, thus posing a challenge in accommodating growing bone tissue. As improving the degradation rate of bioceramics is a key objective in much recent biomaterial research [51], there exists somewhat of a trade-off in achieving both good mechanical performance and bioresorpability. Nevertheless, for the materials fabricated here, toughness values in terms of $K_{IC}$ were higher than the previously reported data for akermanite (1.83 MPa·m$^{1/2}$) [4], bredigite (1.57 MPa·m$^{1/2}$) [5], wollastonite (<1 MPa·m$^{1/2}$) [6],β-TCP (0.46 MPa·m$^{1/2}$) and HAP (0.70 MPa·m$^{1/2}$) [52]. The improvement of scaffold toughness by the addition of copper is consistent with previous reports that have shown that adding trace elements may enhance the mechanical properties of silicate biomaterials [53]. It is worth noting that, as shown in Table 2 in the scaffold materials studied here, improved mechanical performance was achieved alongside a moderate improvement to biodegradability despite enhanced sintering of these materials. Furthermore, all scaffolds showed continuous and consistent drug release behaviors, indicating that copper incorporation into diopside scaffolds did not change release dynamics.

### 4.2. In vitro bioactivity and biodegradability

The formation of a precipitated biomimetic apatite layer is considered to be an essential requirement for osseointegration, which requires an adequate bonding between artificial material and host bone tissue. The results of XRD measurements (Fig. 5b) and SEM observations (Fig. 6) after immersion in the SBF for 28 days demonstrate that xCu-DIO scaffolds have an excellent apatite-forming ability. Moreover, the pH value (Fig. 5a) was maintained in the range of 7.4–7.8, which provides a good environment for cell metabolism and growth [54]. Therefore, the addition of a certain amount of copper is deemed here to be beneficial to the bioactive behavior of diopside bioceramic scaffolds. The formation mechanism of apatite on the surfaces of bioactive ceramics has been extensively studied [55,56]. Based on this mechanism the apatite precipitation process on the xCu-DIO scaffolds studied here is schematically represented in Fig. 11. Briefly, when the scaffolds are immersed in SBF, $Ca^{2+}$ and $Mg^{2+}$ from the diopside lattice are exchanged with protons in SBF solution leading to the formation of silanol (Si-OH), eventually leading to the production of a negatively charged surface with a silica-rich layer (Si-O-). Later, the adsorption of solute ions from SBF induces the nucleation and growth of apatite precipitates [57].

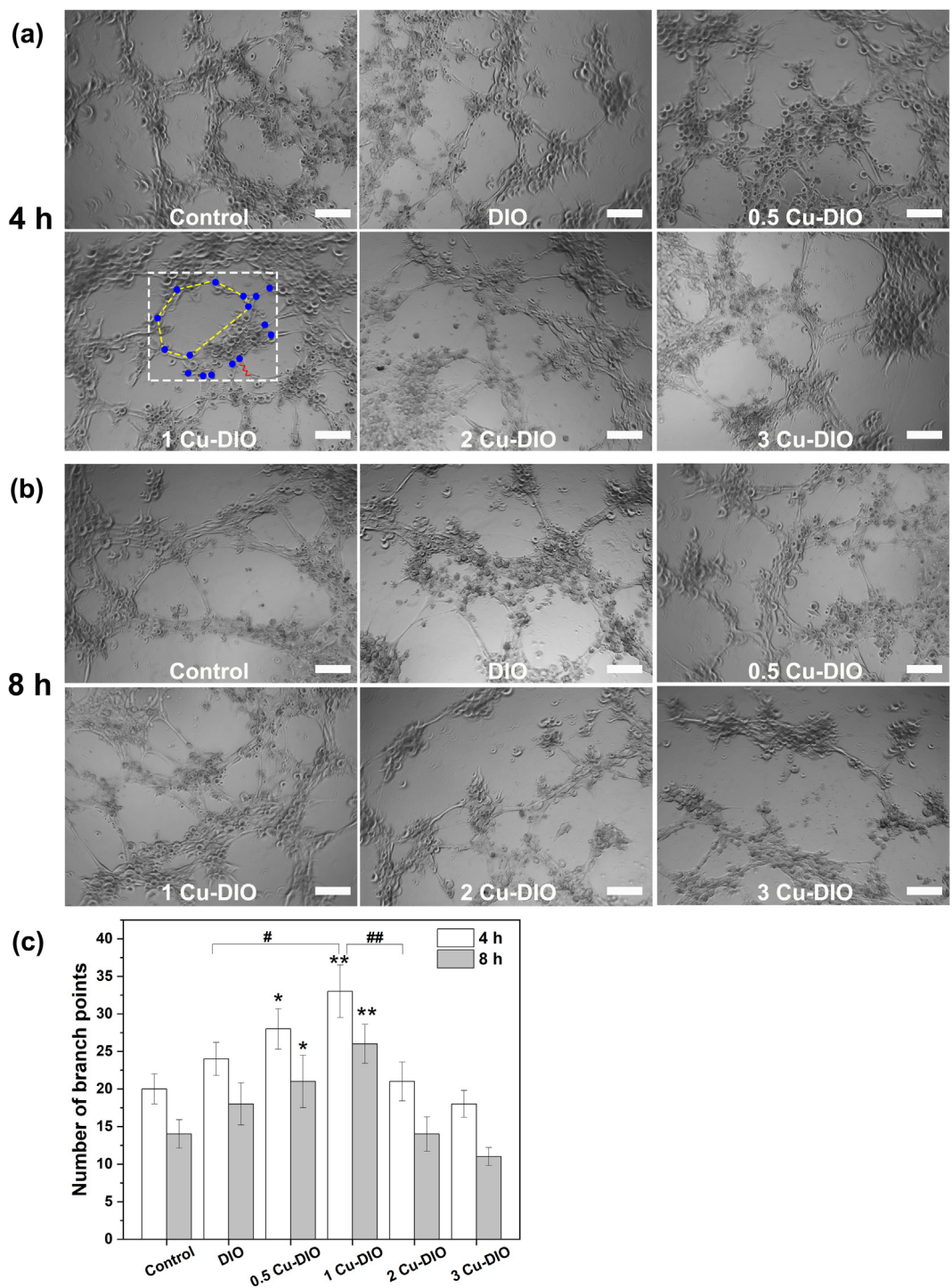

**Fig. 9.** Angiogenic activity of HUVECs cultured in extracts *in vitro*. (a-b) Optical images of HUVECs cultured on matrigel in the presence of Cu-DIO extracts for 4 h and 8 h, where a typical area is boxed in Cu-DIO in (a), blue points indicate branch points, yellow line indicates mesh-like circle, and red lines indicate tube. (Bar = 200 μm). (c) Quantification of the number of branch points (n = 5, *P < 0.05 and **P < 0.01, compared to 4 h or 8 h control group, respectively. #P < 0.05 and ##P < 0.01, as pointed out in the figure).

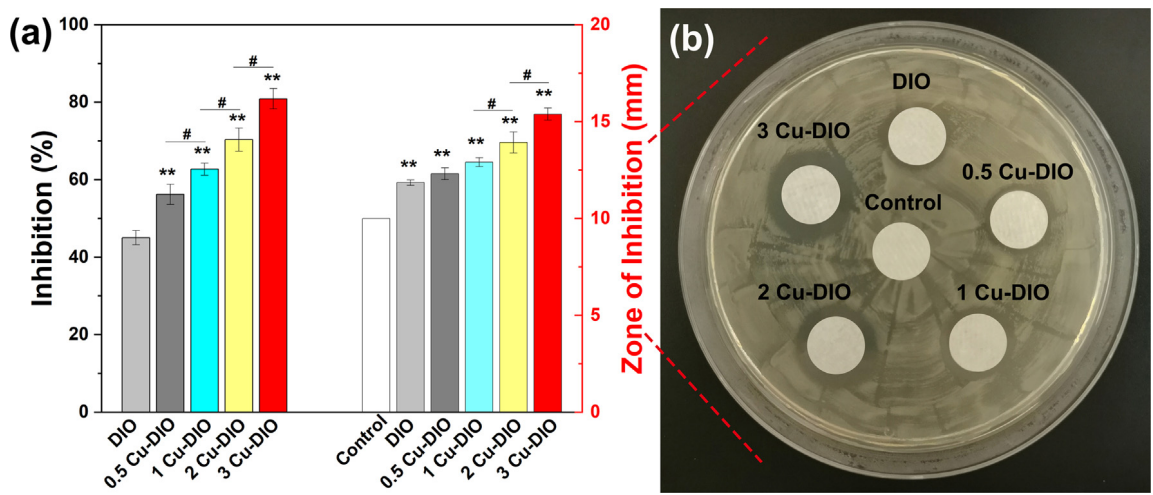

**Fig. 10.** The anti-bacterial performance of xCu-DIO samples by liquid medium microdilution method and agar diffusion assay. (a) Bacterial inhibition rate calculated by microdilution method (left Y-axis) (n = 6) and the quantitative results of inhibition zone (right Y-axis) (*P < 0.05 and **P < 0.01, compared with DIO or control group, respectively). #P < 0.05 and ##P < 0.01, as pointed out in the figure), (b) the typical photographs of cultivated *E. coli* colonies on agar for 24 h in the presence of scaffold extracts on sterilized filter paper. (n = 3).

A significant challenge in the design of bioceramics is the achievement of dissolution rates that match the growth rate of bone. It can be seen from Table 2 that the weight loss increased from 9.3% (DIO) to 11.2% (3 Cu-DIO). Compared with diopside, bioglass has a slightly higher degradation rate. The weight loss of a 45S5 bioglass scaffold following one week immersion in SBF solution was 12% [58]. For pure β-TCP, the weight loss after 6 weeks was 14.21% [43]. The generally sluggish dissolution kinetics of crystalline bioceramics relative to bioglasses are an obstacle towards their development as effective tissue engineering materials in the form of bone implants. The addition of dopants in ceramic lattices tends to impart a modicum of lattice disorder and accelerates chemical reactivity and dissolution. Indeed, an increasing content of copper in the diopside materials fabricated here, as expected brought about moderately enhanced dissolution rates showing the merit of this approach towards enhancing the biodegradation of such silicate bioceramics. The enhanced fracture toughness of the diopside bioceramics studied here, relative to bioglass and other crystalline bioceramics, further implies that lighter, more porous scaffolds can be implemented using this material while offering similar mechanical robustness, thus offsetting possible slower biodegradation kinetics.

### 4.3. Cytocompatibility and in vitro angiogenesis

Bone tissue regeneration involves a complex interplay between osteogenic and angiogenic processes that drive bone growth and tissue repair. For osteoconductive bioactivity cell growth and adhesion are important characteristics and cytotoxic activity must be avoided. In this work, we investigated the effect of copper substitution on the viability of Saos-2 cells, HUVECs and fibroblasts in diopside scaffolds. All three types of cells had similar proliferation trends, the 2 and 3 Cu-DIO materials were clearly toxic to cells, and the effects were dose-dependent. Enhanced adhesion of cells could be observed on 0.5 and 1 Cu-DIO scaffolds. This may be due to the copper ions stimulating the cells to develop more lipid membranes and filopodia that promote cell-substrate and cell–cell interaction. The results of cell proliferation and cell vitality tests were consistent, showing that 0.5 and 1 Cu-DIO have potential osteogenic and angiogenic effects towards Saos-2 cells, HUVECs and fibroblasts (Fig. 7 and Fig. 8). Studies have reported that the cytotoxic effects of copper in biomaterials may be mediated by the formation of OH radicals. Both $Cu^+$ and $Cu^{2+}$ can act as electron donors/acceptors and participate in redox reactions, thereby generating highly reactive oxygen, and then forming OH radicals. As a powerful oxidant, OH radicals will affect biological systems by reacting with proteins, which in turn will cause oxidative damage to these pro-teins [59]. In general, the cytotoxicity caused by $Cu^{2+}$ is proportional to the concentration (when its content exceeds certain limits). Our research shows that when the doping amount of copper at magnesium sites was 1 at.% or lower, the scaffolds were cytocompatible, and for higher copper content, namely 2–3 at.%, they exhibited some degree of cytotoxicity. As the specific threshold concentration of copper cytotoxicity in silicate ceramics depends on its release kinetics, this threshold level would vary between different silicate structures and would further be influenced by the morphology and thermal processing of such ceramics. Angiogenesis is further essential for successful bone repair and copper has been shown to be necessary for activation of endothelial cells in the early stages of angiogenesis [60]. In the present study, we found that Cu-DIO extracts stimulated angiogenesis *in vitro*, as a greater number of branch points were formed in 1 Cu-DIO extract. The mild cytotoxicity of 2–3 Cu-DIO materials limited HUVECs tube formation in these samples.

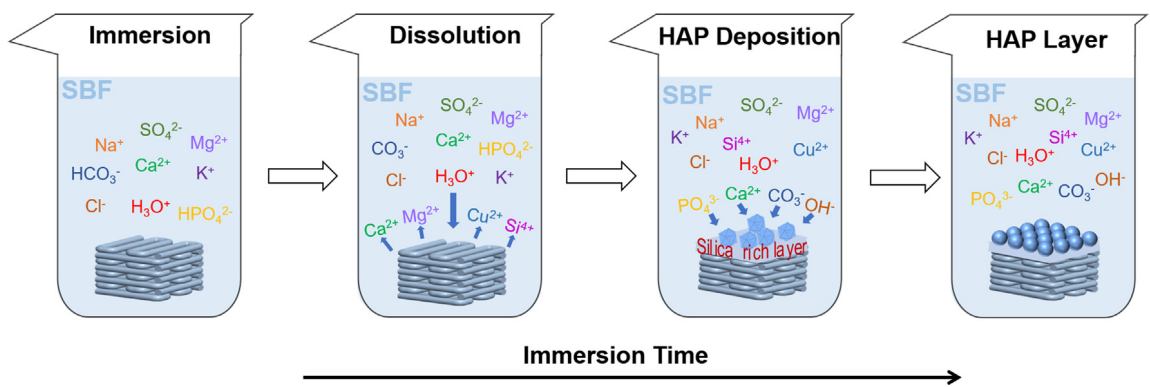

**Fig. 11.** Schematic drawing summarizing the proposed mechanism of apatite layer formation of xCu-DIO scaffolds.

*4.4. Antibacterial activity*

Implant-associated infections remain a thorny problem in tissue reconstruction. The antibacterial effect of copper has been widely reported [10–12]. By incorporating appropriate copper levels in bioceramics, a consistent and safe antibacterial effect may be attained [16]. In this study, copper incorporated into diopside increased the antibacterial effect in a dose-dependent manner (Fig. 10). The antibacterial bioactivity of undoped diopside can be attributed to the increase in pH that occurs as $Ca^{2+}$ and $Mg^{2+}$ from silicate lattice are leached into the aqueous medium [47]. For copper containing bioceramics, the mechanism of antibacterial activity may involve the OH radicals formed by the redox reaction that alternates between $Cu^{+}$ and $Cu^{2+}$, thereby damaging the plasma membrane of *E. coli* in a similar manner to the cytotoxic activity described above [59,61]. We suggest that a synergistic effect of diopside and copper ions brought about the antibacterial effect seen here. It is indicated that all samples have antibacterial properties and can potentially be further used in clinical applications.

In the materials fabricated in this work, we have observed attractive multi-functionalities whereby multiple modes of bioactivity are combined in a scaffold of comparably good compressive strength. The results demonstrate how materials and methods examined in this work can serve to address the structural, mechanical, chemical and biological challenges facing the development of effective bioceramics for bone tissue engineering applications. Due to cytotoxicity exhibited at higher copper contents, 1 at.% copper substituted diopside appears to be the optimal composition from among the materials used in this work.

## 5. Conclusion

In this paper, copper substituted diopside scaffolds were successfully manufactured by robocasting in geometries relevant to bone-tissue engineering applications. The results show that copper doping at up to 3% of magnesium sites had no significant effect on the structure of diopside. The scaffolds had high porosity (~80%), and the mechanical properties were improved by copper addition, exceeding the strength of cancellous bone. xCu-DIO scaffolds showed good osseointegrative apatite formation and sustained drug release abilities. Biodegradation was accelerated with the increase of copper content, and the degraded scaffolds still meeting the strength requirements of cancellous bone implants. Diopside scaffolds with low levels of copper substitution show enhanced proliferation and adhesion of Saos-2 cells, HUVECs and fibroblasts, and further promoted angiogenesis, while cytotoxicity arises at higher copper contents. The bioceramics scaffolds studied here were found to have favorable antibacterial activity, which increased with copper content. As discussed, from this work diopside with 1% of magnesium sites substituted with copper (1 Cu-DIO) would appear to be the optimal composition for a bioactive scaffold material with suitable mechanical properties and bioactivity. The findings of this study present encouraging prospects in relation to the design of multifunctional bioceramics based on copper substituted diopside and demonstrate the need for careful materials design in order to effectively harness the positive effects of copper modification in silicate biomaterials.

## CRediT authorship contribution statement

**Shumin Pang:** Conceptualization, Methodology, Investigation, Data curation, Writing – original draft, Writing – review & editing. **Dongwei Wu:** Conceptualization, Methodology, Investigation, Data curation, Writing – original draft, Writing – review & editing. **Franz Kamutzki:** Investigation, Writing – review & editing. **Jens Kurreck:** Resources, Supervision, Writing – review & editing. **Aleksander Gurlo:** Resources, Supervision, Funding acquisition, Writing – review & editing. **Dorian A.H. Hanaor:** Conceptualization, Supervision, Writing – review & editing.

## Declaration of Competing Interest

The authors declare that they have no known competing financial interests or personal relationships that could have appeared to influence the work reported in this paper.

## Acknowledgement

This work was supported by the China Scholarship Council (CSC, 201906780023 &201906780024) . We acknowledge the support of the German Research Foundation and the Open Access Publication Fund of TU Berlin. We also thank M.Sc. Merle Schmahl from Technische Universität Berlin for conducting the nanoindentation measurement.